# FUSION PHAGE AS A BIOSELECTIVE NANOMATERIAL: EVOLUTION OF THE CONCEPT

*Valery A. Petrenko, Sergey N. Ustinov, and I-Hsuan Chen*

Auburn University, Auburn, AL, U.S.A.

*Corresponding and presenting author: 252 Greene Hall, Department of Pathobiology, Auburn University, Al 36849, U.S.A., petreva@auburn.edu;

**ABSTRACT**

Multibillion-clone landscape phage display libraries were prepared by the fusion of the phage major coat protein pVIII with foreign random peptides. Phage particles and their proteins specific for cancer and bacterial cells were selected from the landscape libraries and exploited as molecular recognition interfaces in detection, gene- and drug-delivery systems. The biorecognition interfaces were obtained by incorporation of the cell-specific phage fusion proteins into liposomes using intrinsic structural duplicity of the proteins. As a paradigm, we incorporated targeted pVIII proteins into commercially available therapeutic liposomes "Doxil", which acquired a new emergent property—ability to bind target receptors. Targeting of the drug was evidenced by fluorescence-activated cell sorting, microarray, optical and electron microscopy. In contrast to a poorly controllable conjugation targeting, the new landscape phage-based approach relies on very powerful and extremely precise mechanisms of selection, biosynthesis and self assembly, in which phages themselves serve as a source of the final product.

## 1. INTRODUCTION

Phage display technology emerged as a synergy of two fundamental concepts: *combinatorial peptide libraries* and *fusion phage*. The first concept replaced the traditional collections of natural or individually synthesized compounds for libraries of peptides obtained in parallel synthesis as grouped mixtures [1, 2] (reviewed in [3]); the second—allowed displaying foreign peptides on the surface of bacterial viruses (bacteriophages) as part of their minor or major coat proteins [4, 5](reviewed in [6, 7]). The merge of these two concepts resulted in development of *phage display libraries*—multibillion clone compositions of self-amplified and self-assembled biological particles [6]. In particular, a paradigm of *landscape phage libraries* evolved, in which the phage is considered not just as a genetic carrier for foreign peptides, like in the traditional phage display approach, but rather as a nanoparticle (nanotube) with emergent physico-chemical characteristics determined by specific phage landscapes formed by thousands of random peptides fused to the major coat protein pVIII [8]. These constructs display the guest peptide on every pVIII subunit increasing the virion's total mass by 10%. Despite of the extra burden, such particles could retain their infectivity and progeny-forming ability.

There is a fast growing interest to the landscape phages as a new type of selectable nanomaterials. The landscape phage can bind organic ligands, proteins and antibodies [8, 9], induce specific immune responses in animals [10-13], or resist stress factors such as chloroform or high temperature [8, 14]. Landscape phages have been shown to serve as substitutes for antibodies against cell-displayed antigens and receptors [15-20], diagnostic probes for bacteria and spores [21-24], gene- and drug-delivery systems [20, 25], and biospecific adsorbents [26]. Phage-derived probes inherit the extreme robustness of wild-type phage [14] and allow fabrication of bioselective materials by self-assemblage of phage or its composites on metal, mineral or plastic surfaces [24, 27-32]. Landscape phages specific for bacterial cells and spores have been exploited as molecular recognition interfaces in detection systems [27, 33].

In this report we first demonstrate the use of the landscape phage as a bioselective interface in drug delivery systems. The concept of using targeted pharmaceutical nanocarriers to enhance the efficiency of anti-cancer drugs has been proven over the past decade both in pharmaceutical research and clinical setting. Examples of a successful realization of this concept are listed in numerous reviews, for example [34-37]. In particular, it is commonly accepted that selectivity of drug delivery systems can be increased by their coupling with peptide and protein ligands targeted to differentially expressed receptors (reviewed in [38-40]). The abundance of these receptors was demonstrated recently by comparative analysis of gene expression in tumor cells and tumor vascular endothelial cells versus adjacent normal tissues [41], and their targeting is turning into a routine procedure in the most advanced laboratories due to the progress in combinatorial chemistry and phage display (a long lists of different target-specific peptides identified by phage display can be found in recent reviews [6, 39, 43-45]. In particular, selection protocols were developed for obtaining the phages and peptides internalizing into cancer cells [19, 46-50], and phage homing at tumors of human patients [51]

A new challenge, within the frame of this concept, is to develop highly selective, stable, active and physiologically acceptable ligands that would navigate the encapsulated drugs to the site of the disease and control unloading of their toxic cargo inside the cancer cells. We have shown earlier that the tumor-specific peptides fused to the major coat protein pVIII can be affinity selected from multibillion clone landscape phage libraries by their ability to bind very specifically cancer cells [16, 19]. Here we show that the target-specific phage can be converted easily into the drug-loaded

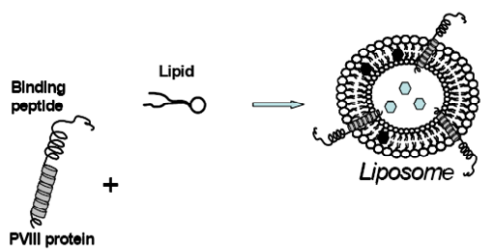

**Fig.1**. Drug-loaded liposome targeted by the pVIII protein. The hydrophobic helix of pVIII spans the lipid layer and binding peptide is displayed on the surface of the carrier particles. The drug molecules are shown as hexagons

vesicles in which the fusion phage proteins span the lipid bilayer of liposomes or micelles and display the target-binding peptides on the surface of the vesicles, as illustrated in **Fig.1**. Thus, the major principle of our approach is that targeted drug-loaded nanoparticles recognize the same receptors, cells, tissues and organs that have been used for selection of the precisely targeted landscape phage.

## 2. LANDSCAPE PHAGE LIBRARIES AS A SOURCE OF BIOSELECTIVE PROBES

The filamentous bacteriophages Ff (fd, f1 and M13) are long, thin viruses, which consist of a single-stranded circular DNA packed in a cylindrical shell composed of the major coat protein pVIII (98% of the total protein mass), and a few copies of the minor coat proteins capping the ends of the phage particle **(Fig. 2)**. Foreign peptides were displayed on the pVIII protein [5] soon after the display on the minor coat protein pIII was pioneered (reviewed in [7]). The pVIII-fusion phages display the guest peptide on every pVIII subunit, increasing the virion's total mass by up to 15%. Such particles were given the name "landscape phage" to emphasize the dramatic change in surface architecture caused by arraying thousands of copies of the guest peptide in a dense, repeating pattern around the tubular capsid, as illustrated by **Fig. 3** [8]. It was shown that the foreign peptides replacing three or four mobile amino acids close to the N-terminus of the wild-type protein pVIII don't disturb considerably the general architecture of virions and don't change the conformation of the fusion proteins in membranes [13, 52, 53]. A *landscape library* is a diverse population of phages, encompassing billions of clones with different surface structures and biophysical properties [7]. Therefore, the landscape phage is unique micro-fibrous material that can be selected in the affinity binding protocol and obtained by a routine and simple microbiological procedure. Binding peptides, 8-9 amino acid long, comprising up to 20% of the phage mass may be easily prepared by cultivation of the infected bacteria and isolation of the secreted phage particles by precipitation. Landscape phages have been shown to serve as substitutes for antibodies against various antigens and receptors, including live cancer and bacterial cells, in gene-delivery vehicles, and as analytical probes in biosensors. Although phages themselves are very attractive bioselective materials, alternative targeted forms containing phage proteins may be more advantageous in medical applications where nanoparticulate materials are required, such as bioselective drug delivery liposomes described below.

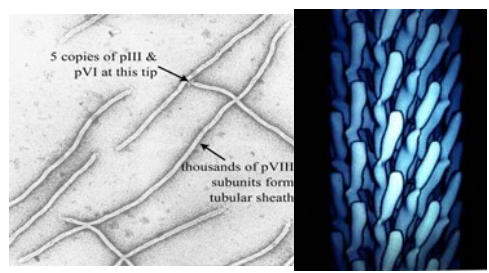

**Fig. 2**. Filamentous phage. Left: electron micrograph. The aminoterminus of pIII proteins are pointed by arrow. Right: Segment of ~1% of phage virion with the array of the pVIII proteins shown as electron densities (Micrograph and model courtesy of Gregory Kishchenko)

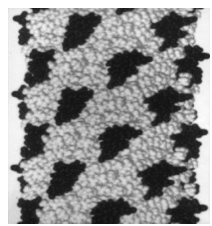

**Fig. 3**. Structure of landscape phage. Foreign peptides are pictured with dark atoms; their overall arrangement corresponds to the model of Marvin, 1994.

## 3. PHAGE PROTEINS IN LIPOSOMES

The ability of the major coat protein pVIII to form micelles and liposomes emerges from its intrinsic function as membrane protein judged by its biological,

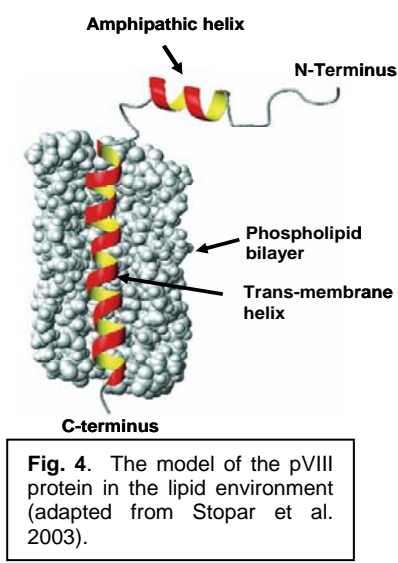

**Fig. 4.** The model of the pVIII protein in the lipid environment (adapted from Stopar et al. 2003).

chemical, and structural properties. During infection of the host *Escherichia coli*, the phage coat is dissolved in the bacterial cytoplasmic membrane, while viral DNA enters the cytoplasm [12]. The protein is synthesized in infected cell as a water-soluble precursor, which contains a leader sequence of 23 residues at its N-terminus. When this protein is inserted into the membrane, the leader sequence is cleaved off by a leader peptidase. Later, during the phage assembly, the processed pVIII proteins are transferred from the membrane into a coat of the emerging phage. Thus, the major coat protein can change its conformation to accommodate to various distinctly different forms of the phage and its precursors: phage filament, intermediate particle (I-form), spheroid (S-form), and membrane-bound form. This structural flexibility of the major coat protein is determined by its unique architecture, which is studied in much detail [17]. In virions, mostly α-helical domain of pVIII are arranged in layers with a 5-fold rotational symmetry and approximate 2-fold screw symmetry around the filament axis, as shown on the **Fig. 3**, right. In opposite, in the membrane-bound form of fd coat protein, the 16-Å-long amphipathic helix (residues 8-18) rests on the membrane surface, while the 35-Å-long trans-membrane (TM) helix (residues 21-45) crosses the membrane at an angle of 26° up to residue Lys40, where the helix tilt changes, as illustrated by **Fig. 4.** The helix tilt accommodates the thickness of the phospholipid bilayer, which is 31 Å for

*E. coli* membrane components. Tyr 21 and Phe 45 at the lipid–water interfaces delimit the TM helix, while a half of N-terminal and the last C-terminal amino acids, including the charged lysine side chains, emerge from the membrane interior. The transmembrane and amphipathic helices are connected by a short turn (Thr 19–Glu 20) that differs from the longer hinge loop (residues 17–26).

## 4. TARGETING OF THE ANTICANCER DRUG DOXIL BY THE STRIPPED LANDSCAPE PHAGE

### 4.1. Stripped phage proteins

In model experiments, landscape phages were converted to a new biorecognition affinity reagent — "stripped phage" [8]. The stripped phage is a composition of disassembled phage coat proteins with dominated (98%) recombinant major coat protein pVIII, which is genetically fused to the foreign target-binding peptides. The stripped phage proteins can form bioselective vesicles decorated by target-binding peptides, which can be used for the targeted drug delivery. In our example, the stripped phages were prepared by treatment of the landscape phages with chloroform (**Fig. 5**) followed by conversion of resulted spheroids into the lipid vesicles by their reconstruction with phospholipids. In preliminary experiments with streptavidin- and bacterial binders it was demonstrated by competition ELISA, acoustic wave sensor and transmission electron microscopy that the stripped phage proteins retain the target-binding properties of the selected phage[8, 54].

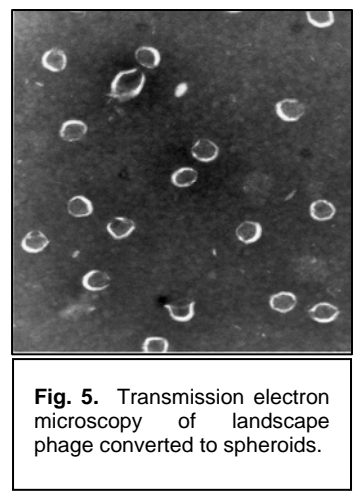

**Fig. 5.** Transmission electron microscopy of landscape phage converted to spheroids.

## 4.2. Fusion of the stripped phage proteins with Doxil

Using intrinsic mechanism of fusion of the phage proteins with lipid membranes, we incorporated streptavidin-targeted proteins into the commercially available Doxil liposomes. The streptavidin-binding landscape phage was affinity selected from 9-mer landscape library. The phage was converted into spheroids with chloroform and incubated with Doxil to allow fusion of the phage proteins with liposome membrane, as illustrated by **Fig. 1**. As a result of the phage fusion, the liposome acquired a new emergent property—ability to bind streptavidin and streptavidin-conjugated fluorescent molecules, as was evidenced by protein microarrays, fluorescent microscopy and fluorescence-activated cell sorting (FACS). The targeted and control liposomes were incubated with streptavidin-coated chips, washed and scanned (**Fig. 6**), or mixed with Texas Red-conjugated streptavidin (TRS), washed and analyzed by fluorescent microscopy and FACS. Complex of the modified Doxil with the target streptavidin demonstrated 50-fold higher fluorescence than pure Doxil and 10-fold higher fluorescence than control Doxil treated with TRS, as registered by FL6 channel specific for fluorescence of the Texas Red label (data are not shown). No significant changes of fluorescent signals were registered in the FL2 channel, more specific for doxorubicin. Complex of the targeted Doxil liposomes with streptavidin-coated gold beads was visualized by transmission electron microscopy (**Fig. 7**).
.

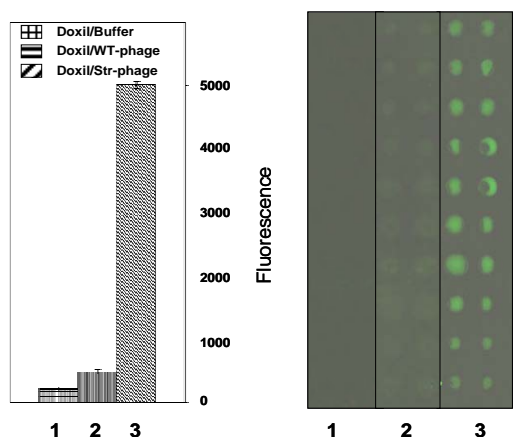

**Fig. 6**. Microarray test for Doxil targeting. Streptavidin-microarrayed slide was treated with unmodified Doxil (1), Doxil modified with wild-type phage (2) and Doxil targeted with streptavidin binding phage (3).

## 5. CONCLUSION. LANDSCAPE PHAGE AS BIOSELECTIVE VEHICLE FOR TARGETED DRUG DELIVER

We proved a novel approach for specific targeting of pharmaceutical nanocarriers through their fusion with stripped proteins of the affinity selected landscape phage. The phage specific for the target organ, tissue or cell is selected from the multibillion landscape phage libraries, and is combined with liposomes exploring intrinsic amphiphilic properties of the phage proteins. As a result, the targeting probe—the target-specific peptide fused to the major coat protein—is exposed on the shell of the drug-loaded vesicle. In contrast to sophisticated and poorly controllable conjugation procedures used for coupling of synthetic peptides and antibodies to the targeted vesicles, the new phage-based concept relies on very powerful and extremely precise mechanisms of *selection, biosynthesis and self assembly*. When landscape phage serve as a reservoir of the targeted membrane proteins, one of the most troublesome steps of the conjugation technology is bypassed. Furthermore, it does not require idiosyncratic reactions with any new shell-decorating polymer or targeting ligand and may be easily adapted to a new liposome or micelle composition and a new addressed target. No reengineering of the selected phage is required at all: the phages themselves serve as the source of the final product—coat protein genetically fused to the targeting peptide. A culture of phage-secreting cells is an efficient, convenient and discontinuous protein production system. They are secreted from the cell nearly free of intracellular components; their further purification could be easily accomplished by simple steps that do not differ from one clone to another. The major coat protein constitutes 98% of the total protein mass of the virion — the purity hardly reachable in normal synthetic and bioengineering procedures. As a normal intestinal parasite, phage itself and its components are not toxic and have been already tested for safety in preclinical and clinical trials [51, 55]. In contrast to immunization procedure, the phage selection protocol may require tiny amounts of a target material (thousands of tumor cells available in biopsy procedure [42] for obtaining the tumor-specific phage ligands, affinity and selectivity of which may be controlled by exploring well developed depletion and affinity maturation procedures.

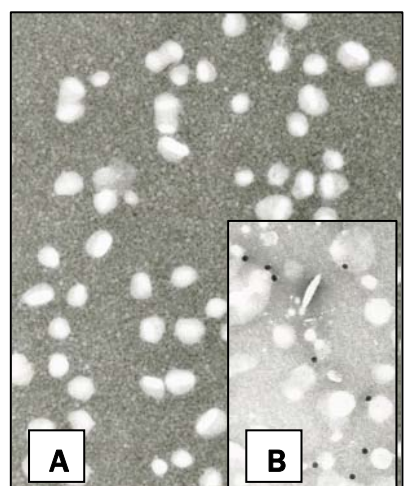

**Fig. 7.** Transmission electron microscopy of original Doxil liposomes (A), and complexes of the targeted Doxil liposomes with streptavidin-coated 20 nm gold beads. An evarage size of unmodified Doxil particals – 80 nm.

## 6. REFERENCES

[1]. Geysen, H.M., R.H. Meloen, and S.J. Barteling, *Use of peptide synthesis to probe viral antigens for epitopes to a resolution of a single amino acid.* Proceedings Of The National Academy Of Sciences Of The United States Of America, 1984. **81**(13): p. 3998-4002.

[2]. Houghten, R.A., *General-Method for the Rapid Solid-Phase Synthesis of Large Numbers of Peptides - Specificity of Antigen-Antibody Interaction at the Level of Individual Amino-Acids.* Proceedings of the National Academy of Sciences of the United States of America, 1985. **82**(15): p. 5131-5135.

[3]. Geysen, H.M., et al., *Combinatorial compound libraries for drug discovery: An ongoing challenge.* Nature Reviews Drug Discovery, 2003. **2**(3): p. 222-230.

[4]. Smith, G.P., *Filamentous fusion phage: novel expression vectors that display cloned antigens on the virion surface.* Science, 1985. **228**(4705): p. 1315-7.

[5]. Ilyichev, A.A., et al., *Construction of M13 viable bacteriophage with the insert of foreign peptides into the major coat protein.* Doklady Biochemistry (Proc.Acad. Sci. Ussr)-Engl.Tr., 1989. **307**: p. 196-198.

[6]. Smith, G.P. and V.A. Petrenko, *Phage Display.* Chemical Reviews, 1997. **97**(2): p. 391-410.

[7]. Petrenko, V.A. and G.P. Smith, *Vectors and Modes of Display*, in *Phage Display in Biotechnology and Drug Discovery*, S.S. Sidhu, Editor. 2005, CRC Pressw, Taylor & Francis Group: Bo Raton, FL, U.S.A. p. 714 pp.

[8]. Petrenko, V.A., et al., *A library of organic landscapes on filamentous phage.* Protein Engineering, 1996. **9**(9): p. 797-801.

[9]. Petrenko, V.A., et al., *Alpha-helically constrained phage display library.* Protein Engineering, 2002. **15**(11): p. 943-950.

[10]. Minenkova, O.O., et al., *Design of specific immunogens using filamentous phage as the carrier.* Gene, 1993. **128**(1): p. 85-88.

[11]. di Marzo Veronese, F., et al., *Structural mimicry and enhanced immunogenicity of peptide epitopes displayed on filamentous bacteriophage. The V3 loop of HIV-1 gp120.* Journal of Molecular Biology, 1994. **243**(2): p. 167-72.

[12]. Perham, R.N., et al., *Engineering a peptide epitope display system on filamentous bacteriophage.* FEMS Microbiology Reviews, 1995. **17**(1-2): p. 25-31.

[13]. De Berardinis, P., et al., *Phage display of peptide epitopes from HIV-1 elicits strong cytolytic responses.* Nature Biotechnology, 2000. **18**(8): p. 873-876.

[14]. Brigati, J. and V. Petrenko, *Thermostability of landscape phage probes.* Anal Bioanal Chem., 2005. **382**(6): p. 1346-50.

[15]. Petrenko, V.A. and G.P. Smith, *Phages from landscape libraries as substitute antibodies.* Protein Engineering, 2000. **13**(8): p. 589-592.

[16]. Romanov, V.I., D.B. Durand, and V.A. Petrenko, *Phage display selection of peptides that affect prostate carcinoma cells attachment and invasion.* Prostate, 2001. **47**(4): p. 239-251.

[17]. Lee, S., M.F. Mesleh, and S.J. Opella, *Structure and dynamics of a membrane protein in micelles from three solution NMR experiments.* Journal of Biomolecular NMR, 2003. **26**(4): p. 327-34.

[18]. Bishop-Hurley, S.L., et al., *Phage-displayed peptides as developmental agonists for Phytophthora capsici zoospores.* Applied and Environmental Microbiology, 2002. **68**(7): p. 3315-3320.

[19]. Samoylova, T.I., et al., *Phage Probes for Malignant Glial Cells.* Molecular Cancer Therapeutic, 2003. **2**: p. 1129-1137.

[20]. Mount, J.D., et al., *Cell Targeted Phagemid Rescued by Pre-Selected Landscape Phage.* Gene, 2004. **In Press**.

[21]. Petrenko, V.A. and V.J. Vodyanoy, *Phage display for detection of biological threat agents.* Journal of Microbiological Methods, 2003. **53**(2): p. 253-62.

[22]. Brigati, J., et al., *Diagnostic probes for Bacillus anthracis spores selected from a landscape phage library.* Clinical Chemistry, 2004. **In Press**.

[23]. Petrenko, V.A. and I.B. Sorokulova, *Detection of biological threats. A challenge for directed molecular evolution.* Journal of Microbiological Methods, 2004. **58**(2): p. 147-168.

[24]. Sorokulova, I.B., et al., *Landscape phage probes for Salmonella typhimurium.* J. Microbiol. Methods., 2005.

[25]. Petrenko, V., *Drug Delivery Nanocarriers Targeted by Landscape Phage.* 2006: U.S. Utility Patent Application.

[26]. Samoylova, T.I., et al., *Phage matrix for isolation of glioma cell-membrane proteins.* Biotechniques, 2004. **37**(2): p. 254-260.

[27]. Olsen, E.V., et al., *Affinity-selected filamentous bacteriophage as a probe for acoustic wave biodetectors of Salmonella typhimurium.* Biosens Bioelectron, 2006. **21**(8): p. 1434-42.

[28]. Nanduri, V., et al., *Phage as a molecular recognition element in biosensors immobilized by physical adsorption.* Biosens Bioelectron, 2006.

[29]. Flynn, C.E., et al., *Viruses as vehicles for growth, organization and assembly of materials.* Acta Materialia, 2003. **51**(19): p. 5867-5880.

[30]. Flynn, C.E., et al., *Synthesis and organization of nanoscale II-VI semiconductor materials using evolved peptide specificity and viral capsid assembly.* Journal of Materials Chemistry, 2003. **13**(10): p. 2414-2421.

[31]. Reiss, B.D., et al., *Biological routes to metal alloy ferromagnetic nanostructures.* Nano Letters, 2004. **4**(6): p. 1127-1132.

[32]. Mao, C.B., et al., *Virus-based toolkit for the directed synthesis of magnetic and semiconducting nanowires.* Science, 2004. **303**(5655): p. 213-217.

[33]. Wan, J., et al. *Phage-based magnetostrictive-acoustic microbiosensors for detecting Bacillus anthracis spores.* in *Defense and Security Symposium.* 2006. Gaylord Palms Resort and Convention Center. Orlando (Kissimmee), Florida USA.

[34]. Noble, C.O., et al., *Development of ligand-targeted liposomes for cancer therapy.* Expert Opin Ther Targets, 2004. **8**(4): p. 335-53.

[35]. Torchilin, V.P., *Drug targeting.* Eur J Pharm Sci, 2000. **11 Suppl 2**: p. S81-91.

[36]. Vasir, J.K. and V. Labhasetwar, *Targeted drug delivery in cancer therapy.* Technol Cancer Res Treat, 2005. **4**(4): p. 363-74.

[37]. Everts, M., *Targeted therapies directed to tumor-associated antigens.* Drugs of the future, 2005. **30**(10): p. 1067-1076.

[38]. Krumpe, L.R.H. and T. Mori, *The use of phage-displayed peptide libraries to develop tumor-targeting drugs.* International Journal of Peptide Research and Therapeutics, 2006. **12**(1): p. 79-91.

[39]. Shadidi, M. and M. Sioud, *Selective targeting of cancer cells using synthetic peptides.* Drug Resist Updat, 2003. **6**(6): p. 363-71.

[40]. Mori, T., *Cancer-specific ligands identified from screening of peptide-display libraries.* Current Pharmaceutical Design, 2004. **10**: p. 2335-2343.

[41]. Nanda, A. and B. St Croix, *Tumor endothelial markers: new targets for cancer therapy.* Curr Opin Oncol, 2004. **16**(1): p. 44-9.

[42]. Shukla, G.S. and D.N. Krag, *Phage display selection for cell-specific ligands: development of a screening procedure suitable for small tumor specimens.* J Drug Target, 2005. **13**(1): p. 7-18.

[43]. Nilsson, F., et al., *The use of phage display for the development of tumour targeting agents.* Advanced Drug Delivery Reviews, 2000. **43**: p. 165-196.

[44]. Romanov, V.I., *Phage display selection and evaluation of cancer drug targets.* Curr Cancer Drug Targets, 2003. **3**(2): p. 119-29.

[45]. Aina, O.H., et al., *Therapeutic cancer targeting peptides.* Biopolymers, 2002. **66**: p. 184-199.

[46]. Hong, F.D. and G.L. Clayman, *Isolation of a peptide for targeted drug delivery into human head and neck solid tumors.* Cancer Res, 2000. **60**(23): p. 6551-6.

[47]. Ivanenkov, V., F. Felici, and A.G. Menon, *Uptake and intracellular fate of phage display vectors in mammalian cells.* Biochim Biophys Acta, 1999. **1448**(3): p. 450-62.

[48]. Ivanenkov, V.V., F. Felici, and A.G. Menon, *Targeted delivery of multivalent phage display vectors into mammalian cells.* Biochim Biophys Acta, 1999. **1448**(3): p. 463-72.

[49]. Ivanenkov, V.V. and A.G. Menon, *Peptide-mediated transcytosis of phage display vectors in MDCK cells.* Biochem Biophys Res Commun, 2000. **276**(1): p. 251-7.

[50]. Zhang, J., H. Spring, and M. Schwab, *Neuroblastoma tumor cell-binding peptides identified through random peptide phage display.* Cancer Lett, 2001. **171**(2): p. 153-64.

[51]. Krag, D.N., et al., *Selection of tumor-binding ligands in cancer patients with phage isplay libraries.* Cancer Research, 2006. **66**(15): p. 7724-7733.

[52]. Jelinek, R., et al., *NMR structure of the principal neutralizing determinant of HIV-1 displayed in filamentous bacteriophage coat protein.* Journal of Molecular Biology, 1997. **266**(4): p. 649-55.

[53]. Monette, M., et al., *Structure of a malaria parasite antigenic determinant displayed on filamentous bacteriophage determined by NMR spectroscopy: implications for the structure of continuous peptide epitopes of proteins.* Protein Sci, 2001. **10**(6): p. 1150-9.

[54]. Olsen, E., et al., *Phage Fusion Proteins as Bioselective Receptors for Piezoelectric Sensors.* "ECS Transactions - Denver" Volume 2,"Biosensor Systems", 2006. **In Press**.

[55]. Krag, D.N., et al., *Phage-displayed random peptide libraries in mice: toxicity after serial panning.* Cancer Chemother Pharmacol, 2002. **50**(4): p. 325-32.

ACKNOWLEDGMENT

We are grateful to Dr. M.A. Toivio-Kinnucan (College of Veterinary Medicine, Auburn University) for excellent electron microscopic support.